\documentclass{SciPost}

\usepackage{bm}
\usepackage{subcaption}
\usepackage{longtable}
\usepackage{booktabs}
\usepackage{array}
\usepackage{float}
\hypersetup{
    colorlinks,
    linkcolor={red!50!black},
    citecolor={blue!50!black},
    urlcolor={blue!80!black},
}

\usepackage[bitstream-charter]{mathdesign}
\DeclareSymbolFont{usualmathcal}{OMS}{cmsy}{m}{n}
\DeclareSymbolFontAlphabet{\mathcal}{usualmathcal}

\fancypagestyle{SPstyle}{
\fancyhf{}
\lhead{\colorbox{scipostblue}{\bf \color{white} ~SciPost Physics }}
\rhead{{\bf \color{scipostdeepblue} ~Submission }}

\fancyfoot[C]{\textbf{\thepage}}
}

\begin{document}

\pagestyle{SPstyle}

\begin{center}{\Large \textbf{\color{scipostdeepblue}{
SuperJEPA: Pretraining with a Supervision-Augmented Joint-Embedding Predictive Architecture for High Energy Physics
}}}\end{center}

\begin{center}\textbf{
Dylan Sheldon Rankin\textsuperscript{1$\star$} and
Ho Fung Tsoi\textsuperscript{1}
}\end{center}

\begin{center}
{\bf 1} Department of Physics and Astronomy, University of Pennsylvania, Philadelphia, PA, USA
\\[\baselineskip]
$\star$ {\small dsrankin at sas.upenn.edu}
\end{center}

\section*{\color{scipostdeepblue}{Abstract}}
\textbf{\boldmath{%
Foundation models often obtain powerful results through very large unlabeled training datasets that dwarf available labeled datasets in size, but high energy physics already provides large samples of high-fidelity labeled simulation. 
We therefore ask whether self-supervised objectives improve over supervised pretraining alone when large labeled samples already exist. 
We introduce \textit{SuperJEPA}, a joint-embedding predictive architecture augmented with a supervised objective, and find improved downstream jet-tagging performance across multiple tasks, model sizes, and pretraining dataset sizes. 
The gains persist for both frozen and fine-tuned encoders, motivating predictive self-supervision in future large-scale particle-physics foundation models.
}}

\vspace{\baselineskip}

\noindent\textcolor{white!90!black}{%
\fbox{\parbox{0.975\linewidth}{%
\textcolor{white!40!black}{\begin{tabular}{lr}%
  \begin{minipage}{0.6\textwidth}%
    {\small Copyright attribution to authors. \newline
    This work is a submission to SciPost Physics. \newline
    License information to appear upon publication. \newline
    Publication information to appear upon publication.}
  \end{minipage} & \begin{minipage}{0.4\textwidth}
    {\small Received Date \newline Accepted Date \newline Published Date}%
  \end{minipage}
\end{tabular}}
}}
}


\vspace{10pt}
\noindent\rule{\textwidth}{1pt}
\tableofcontents
\noindent\rule{\textwidth}{1pt}
\vspace{10pt}

\section{Introduction}
\label{sec:introduction}

Foundation models for natural language and computer vision have revolutionized industry and research over the past few years~\cite{Bommasani:FM,BERT,GPT3,CLIP,MAE,DINOv2}.
These models are typically very large, and after a pretraining phase can be adapted to many downstream tasks~\cite{Bommasani:FM}.
Frontier large-language model (LLM)-based foundation models are now capable of not just simple programming or chatting, but multi-step reasoning on complex topics~\cite{GPT4,DeepSeekR1}.
Although these models have their own proprietary training schemes, their capabilities are built on a crucial initial phase in which they learn to predict the next word in sentences, so-called ``next token prediction''~\cite{GPT3}.
This pretraining scheme allows the models to learn the semantic structure of language without requiring any labels about the topics in a sentence or meanings of phrases.
Methods like this, in which an algorithm is trained with a task constructed from the data itself, are called self-supervised learning (SSL)
, and their use has allowed models to be trained on vast volumes of data that could never have been labeled directly by hand.

In contrast, high-energy physics (HEP), in particular collider physics, has long relied on large quantities of highly accurate simulation
to train powerful AI/ML algorithms
for dedicated tasks like signal classification~\cite{ParticleNet,ParT}, energy and mass regression~\cite{CMS:bRegression,CMS:PNetMassReg}, particle identification~\cite{ATLAS:FTag,ATLAS:GN2,CMS:BTV,DeepJet,ATLAS:EGamma,CMS:EGamma}, and reconstruction~\cite{MLPF}.
The ability to generate labeled simulation has meant that collider physics is not as reliant on SSL techniques compared to other fields.
In spite of this, SSL and pretraining more generally have become popular topics of research as so-called foundation models in other disciplines have made major
strides.
However, previous research in collider physics and jet tagging has struggled to demonstrate significant performance gains for SSL training schemes when large, accurate, labeled datasets are still present~\cite{JJEPA,JPJEPA,Elsharkawy:objectives}.
Even when tested on datasets outside of their training corpus, supervised methods are still able to ultimately enable models to learn powerful, general, performant, representations of data~\cite{OmniLearn,OmniLearned,Sophon,Elsharkawy:objectives}.

In this work we present a method, based on a joint embedding-predictive architecture (JEPA)~\cite{IJEPA}, that combines both SSL and supervised training and demonstrates improved performance over a supervised baseline across a range of jet-tagging tasks and multiple model and dataset sizes.
The manuscript is organized as follows.
We describe the related work in Sec.~\ref{sec:related}, and the datasets and method in Sec.~\ref{sec:datasets} and Sec.~\ref{sec:method}, respectively.
Results are then presented in Sec.~\ref{sec:result} and discussed in Sec.~\ref{sec:discussion}.
\section{Related Work}
\label{sec:related}

Foundation models have attracted significant interest in high-energy physics.
Existing approaches differ in the level of the data they operate on, in whether pretraining uses labels, and in what the model is trained to predict.

\paragraph{Detector-level foundation models.}
Several works pretrain directly on low-level detector data.
FM4NPP~\cite{FM4NPP} trains a state-space model on simulated time-projection-chamber (TPC) spacepoints by predicting the coordinates of nearby hits, and observes power-law scaling of downstream performance with model sizes up to about $10^8$ parameters.
Panda~V2~\cite{PandaDiplomacy} applies DINO-style self-distillation to particle-imaging detectors, including liquid-argon and gaseous TPCs and water
Cherenkov detectors, and matches specialized baselines with only a thousand labeled events.
These models target reconstruction tasks rather than jet-level representations, but they show that self-supervised pretraining can transfer across tasks within a detector.

\paragraph{Supervised and generative jet foundation models.}
OmniLearn~\cite{OmniLearn,OmniLearn2} and its successor OmniLearned~\cite{OmniLearned} pretrain a particle-based transformer~\cite{ParT} with a combination of supervised classification and diffusion or flow-matching
generation, and the latter scales to more than $10^9$ jets drawn from several datasets.
Sophon~\cite{Sophon} similarly uses large-scale supervised pretraining to accelerate resonance searches.
A recent systematic comparison within the OmniLearned framework~\cite{Elsharkawy:objectives} finds that supervised pretraining performs best when downstream labels are plentiful, that masked-particle objectives help most when labels are scarce, and that generative pretraining is needed for generative downstream tasks.
Supervised pretraining requires labeled simulation, however, so the learned representation is shaped by the class definitions of the pretraining dataset.

\paragraph{Masked and autoregressive pretraining.}
Masked Particle Modeling (MPM)~\cite{MPM} adapts BERT-style pretraining to unordered particle sets by masking particles and predicting discrete tokens from a pretrained encoder.
Its successor~\cite{MPMv2} shows that the tokenizer is unnecessary and that reconstructing the masked particles with a continuous generative head performs better.
OmniJet-$\alpha$~\cite{OmniJetAlpha} instead pretrains a GPT-style model to predict the next token of a tokenized jet, and demonstrates transfer from generation to classification.
Follow-up work improves this pretraining by combining it with masked modeling~\cite{OmniJetAlpha:nexttoken}, applies the model to hadronic tau reconstruction~\cite{Tani:taus}, and pretrains it on jets from CMS Open Data~\cite{AspenOpenJets}.
All of these objectives reconstruct the input, either explicitly or as discrete tokens, so model capacity is spent on predicting low-level particle features.

\paragraph{Contrastive and self-distillation methods.}
JetCLR~\cite{JetCLR} learns jet representations through contrastive methods, using augmentations that encode physical symmetries and infrared and collinear safety.
Re-simulation-based self-supervised learning~\cite{RS3L,Rieck:SSL} instead builds positive pairs by re-running stages of the event simulation, so that invariances are defined by the physics of showering and hadronization rather than by hand-designed augmentations, while MACK~\cite{MACK} uses contrastive pretraining to reduce sensitivity to simulation mismodeling.
jBOT~\cite{jBOT} combines particle-level and jet-level self-distillation, and finds that semantically meaningful clustering of jet classes emerges without labels.
These methods avoid representation collapse through negative pairs, teacher networks, or explicit variance and covariance penalties on the embeddings.

\paragraph{Joint-embedding predictive architectures.}
Joint-embedding predictive architectures (JEPAs) predict the representation of a masked region from the representation of the visible context, rather than reconstructing the masked input~\cite{IJEPA}.
Unlike contrastive approaches, JEPAs do not require the construction of pairs through hand-designed transformations such as rotations or smearing, although they do still require a choice of masking scheme.
LeJEPA~\cite{LeJEPA} replaces the exponential-moving-average teacher and stop-gradient required in earlier JEPAs with SIGReg, a regularizer that drives the embedding distribution toward an isotropic Gaussian, and so prevents collapse with a single, fully trainable encoder.
Three works have applied JEPAs to jets.
J-JEPA~\cite{JJEPA} uses Cambridge/Aachen subjets as context and target units and predicts masked subjet representations with an exponential moving average (EMA) target encoder, but is pretrained on about $10^6$ top-quark and QCD jets and shows gains only in the low-label regime.
HEP-JEPA~\cite{HEPJEPA} uses Point-MAE-style patches with contiguous target blocks and an EMA teacher, pretrained on the full JetClass dataset.
JetParticle-JEPA~\cite{JPJEPA} operates directly on particles with a Particle Transformer encoder and predictor.
It masks a contiguous block of particles ordered by nearest-neighbor sequencing and regularizes an EMA-teacher setup with a KoLeo term~\cite{sablayrolles2019spreading}, while SIGReg is examined only as an alternative.
JetParticle-JEPA also applies feature dropout during pretraining to improve robustness to missing particle features, a strategy related to the feature-group gating used here and in OmniLearned~\cite{OmniLearned}.

\paragraph{This work.}
Our approach differs from these JEPAs most notably in that it uses the teacher-free LeJEPA formulation for the first time in HEP.
This means that both the context and target representations come from one single encoder, and collapse is prevented only by SIGReg applied to within-jet residuals of the particle tokens.
We encourage the encoding of meaningful information by the encoder using a predictor which outputs an unordered set of particle representations from learned queries without access to target positions.
The predictor receives only the context representations and a set of learned queries, and is given no information about the positions, features or number of the target constituents as a hint.
All information about the hidden region must therefore be inferred from the context, which is intended to encourage the encoder to represent the structure of the jet rather than local particle features.
The predicted set is compared with the target set through an optimal-transport matching rather than a fixed particle-to-particle correspondence.
The loss is therefore independent of the ordering of the target constituents, and the predictor needs neither positional mask tokens nor the target coordinates.
\section{Datasets}
\label{sec:datasets}

Our work focuses on the task of jet tagging in collider physics, where the task is to 
classify the initiating particle/process from a collection of final state particles that would be measured in a detector, called a jet.
Different original particles result in jets with various distinguishing features such as the number of clusters of particles, or subjets, or the presence of particles that indicate the presence of intermediate states with measurable lifetime.
We use three simulated datasets of varying size and features: JetClass~\cite{ParT}, JetNet~\cite{JetNet}, and Top Tagging~\cite{TopLandscape}.

\subsection{JetClass}

The JetClass dataset consists of 12.5M jets each from 10 different classes ($H\rightarrow b\bar{b}$, $H\rightarrow c\bar{c}$, $H\rightarrow gg$, $H\rightarrow WW^*\rightarrow 4q$, $H\rightarrow WW^*\rightarrow \ell\nu q\bar{q}$, $t\rightarrow bqq$, $t\rightarrow b\ell\nu$, $W\rightarrow qq'$, $Z\rightarrow q\bar{q}$, $Z\rightarrow\nu\bar{\nu}$) for a total of 125M jets.
The generated jets have transverse momentum $p_T>500$~GeV.
Up to 128 particle constituents are kept for each jet.
There are 12 features for each constituent.
They are $\log p_T$ and $\log E$ of the constituent, its position relative to the jet axis $(\Delta\eta, \Delta\phi)$, and the angular distance $\Delta R = \sqrt{\Delta\eta^2 + \Delta\phi^2}$. 
Next are the transverse and longitudinal impact parameters, given as $\tanh d_0$ and $\tanh d_z$. 
Last are five one-hot particle-identification flags: photon, muon, electron, charged hadron and neutral hadron (\texttt{isPhoton}, \texttt{isMuon}, \texttt{isElectron}, \texttt{isChargedHad}, \texttt{isNeutralHad}).
We use JetClass to construct the pretraining dataset for all models.
The entire dataset is split such that each class has 10M/500k/2M events reserved for training/validation/testing, with separate files for each set.

\subsection{JetNet}

The JetNet dataset consists of approximately 170k jets each from 5 different classes ($q$, $g$, $W$, $Z$, $t$) for a total of 880k jets.
A rough mapping of JetNet classes to JetClass classes is that $q$ and $g$ make up $Z\rightarrow\nu\bar{\nu}$, and $W$, $Z$, and $t$ are equivalent to $W\rightarrow qq'$, $Z\rightarrow q\bar{q}$, and $t\rightarrow bqq$, respectively.
The generated jets have transverse momentum $p_T\sim 1$~TeV.
Up to 30 particles are kept for each jet, and the particle features are only $\log p_T$ of the constituent, its position relative to the jet axis $(\Delta\eta, \Delta\phi)$, and the angular distance $\Delta R = \sqrt{\Delta\eta^2 + \Delta\phi^2}$, totaling 4 features.
The data is split such that each class has 50k/20k/100k for training/validation/testing, with separate files for each set.

\subsection{Top Tagging}

The Top Tagging dataset consists of approximately 1M jets each from 2 different classes (\textit{QCD} and \textit{Top}) for a total of 2M jets.
A rough mapping of JetNet classes to JetClass classes is that \textit{QCD} and \textit{Top} are equivalent to $Z\rightarrow\nu\bar{\nu}$ and $t\rightarrow bqq$, respectively.
The generated jets have transverse momentum 550 GeV$<p_T<650$~GeV.
Up to 200 particles are kept for each jet, and the particle features are only $\log p_T$ and $\log E$ of the constituent, its position relative to the jet axis $(\Delta\eta, \Delta\phi)$, and the angular distance $\Delta R = \sqrt{\Delta\eta^2 + \Delta\phi^2}$, totaling 5 features.
The data is split 60/20/20 for training/validation/testing, with separate files for each set.

\section{Method}
\label{sec:method}

\begin{figure}
    \centering
    \includegraphics[width=1\linewidth]{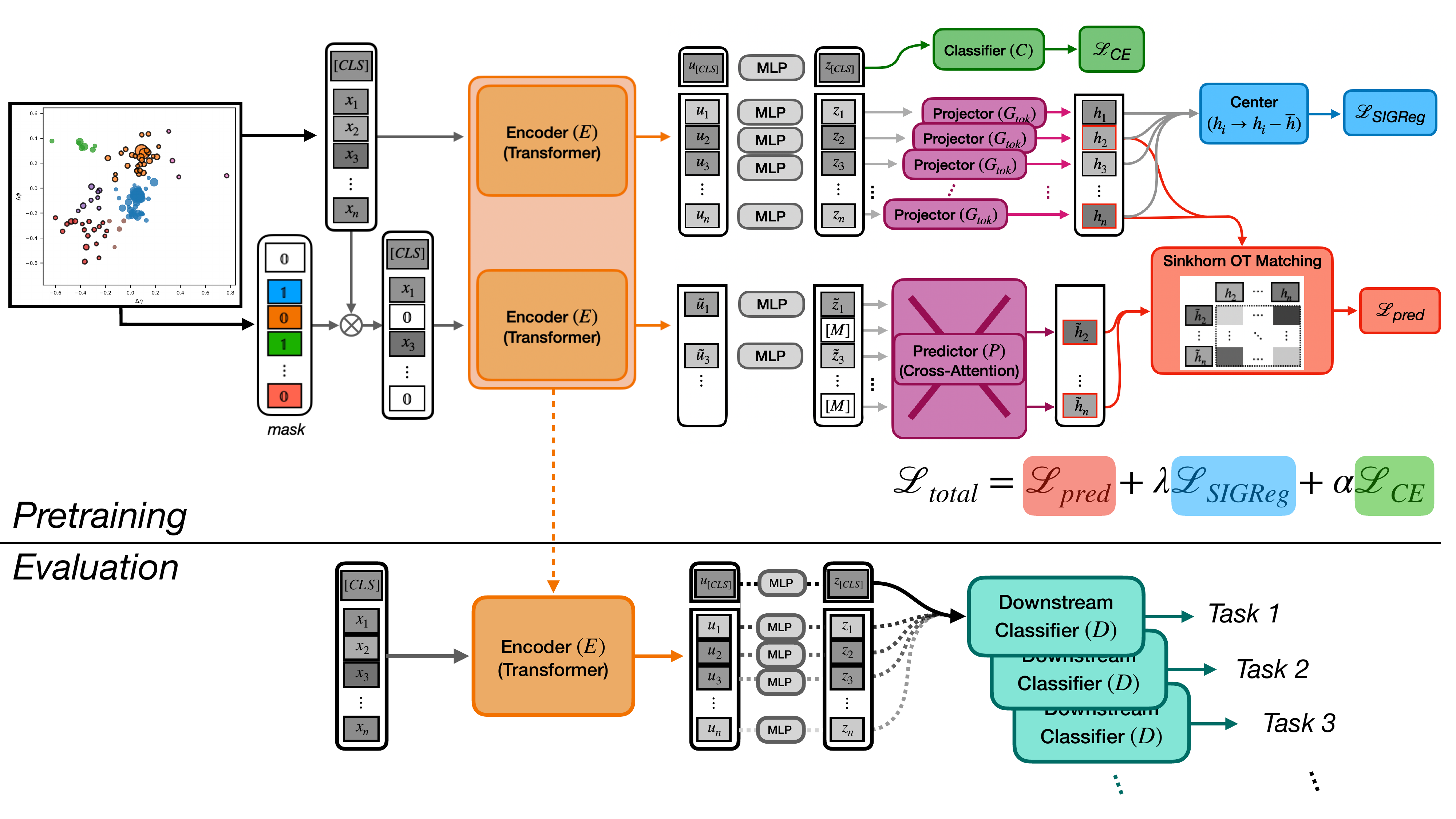}
    \caption{Schematic of the training method and terms that contribute to the loss function.}
    \label{fig:schematic}
\end{figure}

A schematic of SuperJEPA is shown in Fig.~\ref{fig:schematic}.
The method is divided into multiple stages.
A preprocessing stage groups the particles in each jet into subjets that are then used to define ``target’’ and ``context’’ regions.
Two views of the same original jet are then processed through the encoder; in one view the full set of particle tokens is used along with a learnable [CLS] token as input, while in the other only the particles tokens in the ``context’’ are used as input.
The output of the encoder in both cases is a contextual representation for each input token.
The representation is encouraged to have diversity through a SIGReg~\cite{LeJEPA} loss and encouraged to encode meaningful information by forcing the tokens from the ``context’’ view to predict the particle tokens from the ``target’’.
Finally, the [CLS] token is encouraged to encode the relevant high-level class information by its use in a standard cross-entropy classification loss.
The encoder can then be applied in downstream tasks, either directly with its weights frozen or with the weights left tunable.
We discuss these stages in more detail below.
All models are implemented in TensorFlow~\cite{TensorFlow} using the Keras~3 API~\cite{Keras}, and the Cambridge/Aachen reclustering used to define subjets is performed with FastJet~\cite{FastJet}.

\subsection{Preprocessing}

For each jet, we partition the constituents into distinct subjets.
Each jet's constituents are reclustered once with the Cambridge/Aachen (C/A) algorithm using radius $R = 2.0$.
This radius is much larger than the $R = 0.8$ of the JetClass jets, so the reclustering keeps every constituent inside one tree and none is split off as a separate jet.
For clustering, each constituent is treated as a massless four-vector built from its transverse momentum $p_T$ and its jet-relative coordinates $(\Delta\eta, \Delta\phi)$.
The exclusive C/A clustering is stopped at $K_{\max} = 8$ subjets, or at the number of constituents if that is smaller.
Any subjet with fewer than three constituents is then merged into the remaining subjet closest to it in $(\eta, \phi)$.
Merges are repeated one at a time, smallest subjet first, until every subjet has at least three constituents or only one subjet is left.
The resulting $K_{\rm eff} \le K_{\max}$ subjets assign every valid constituent to exactly one subjet, with no particle in two subjets, and are ordered by descending $p_T$.
The assignment is computed once per jet, stored, and reused in every epoch and every view.
The variation between views therefore comes only from which subjets are selected, not from reclustering.

For each jet, one target region $T$ is formed as the union of a subset of these subjets, chosen to contain a set fraction of the jet's transverse momentum.
A target fraction $f^\ast$ is chosen uniformly between 0.10 and 0.35.
Next, we randomly select eight values of $k$ between 1 and $K_{\rm eff}$ - 1  and compute the $p_T$ fraction every combination of $k$ subjets, $f_k = \sum_{s \le k} p_T^{(s)} / \sum_{s} p_T^{(s)}$.
For each of the eight values of $k$, the set whose $f_k$ is closest to $f^\ast$ is selected as the target.
The context is the complement of the target, $C = X \setminus T$,
and the upper bound of $K_{\rm eff} - 1$ ensures that the context is never empty.
Jets with only one subjet cannot be partitioned this way and are excluded from use in the prediction loss (described below).

\subsection{Pretraining}

The model has four main trainable components: a Transformer encoder $E$, a token projector $G_{\rm tok}$, a cross-attention predictor $P$, and a linear classifier $C$.
The same encoder, with shared weights, is applied to the full jet and to the masked context.
Unlike previous usages of JEPA in HEP, we do not use a slowly updated (EMA) copy of the encoder, nor a stop-gradient for training.
Gradients from the prediction loss flow into both the context branch and the target branch.

\subsubsection{Encoder}

The encoder takes each jet as a set of constituents, $X = (x_1, \dots, x_N)$, where each $x_i \in \mathbb{R}^{12}$ holds the features described in Sec.~\ref{sec:datasets}.
In order to allow the use of a model trained with the larger feature-set in JetClass to be applied to JetNet and Top Tagging datasets with fewer features per particle, we employ the scheme from Ref.~\cite{OmniLearn}.
In summary, we apply a trainable embedding layer separately to three feature groups: ($\log p_T, \Delta\eta, \Delta\phi, \Delta R$), ($\log E$), and ($\tanh d_0$, $\tanh d_z$, isPhoton, isMuon, isElectron, isChargedHad, isNeutralHad).
The three embeddings are summed to produce the total $d$-dimensional per-particle embedding that is used as input for our models.
In the case that a dataset lacks a given group, we set that embedding to the null vector.
During training, a special dropout layer randomly masks feature groups per jet per batch to force models to learn embeddings that can be successfully applied without the two groups that may be missing in other datasets.
We set the likelihood of dropping just the final group (Top Tagging-like) to 15\% and the likelihood of dropping the final two groups (JetNet-like) to 5\%.

A learned [CLS] token is prepended to the sequence for the full jet, which then passes through $n$ pre-layer-normalization Transformer blocks.
Each block has $m$-headed self-attention followed by a feed-forward network with hidden width $4d$ and ReLU activation, each wrapped in a residual connection.
No positional encoding is used, so the encoder is permutation-equivariant over constituents, and no dropout is used to prevent the later SuperJEPA components from needing to adapt to this dropout.
A final layer normalization is followed by a two-layer position-wise MLP, $d \to d \to d$ with a ReLU between, applied to every position.
This yields a jet-level embedding $z_{[CLS]}$ and per-particle embeddings $z_1, \dots, z_N$, all of dimension $d$.
We also retain the pre-MLP representations, denoted as $u_{[CLS]}$ and $u_i$, which can be used in downstream tasks.
In particular we find that the $u_i$ are more useful than the $z_i$, presumably as a result of their distance from the SIGReg loss.

During pretraining the encoder is applied twice to each jet.
The full jet $X$ produces the embeddings $z_{[CLS]}$ and $z_i$.
The context $C = X \setminus T$ is the full jet with every target constituent masked, and produces the context embeddings $\tilde{z}_i$.
The $\tilde{z}_{[CLS]}$ is technically produced but is never used; that is, the context view encoder pass is still able to use the CLS token for storing global information about the input set.

\subsubsection{Projector}

The token projector $G_{\rm tok}$ is a two-layer MLP, $d \to 2d \to d$ with a ReLU between.
It is applied independently to each full-jet particle embedding, giving $h_i = G_{\rm tok}(z_i)$.
The projected tokens serve two purposes.
First, the projections $\lbrace h_i : i \in T \rbrace$ of the target constituents are the targets for the predictor.
Second, the projections are centered within each jet,
\begin{equation}
r_i = h_i - \bar{h}, \qquad \bar{h} = \frac{1}{N_{\rm valid}} \sum_{j} h_j,
\end{equation}
where the sum runs over the valid constituents of the jet.
SIGReg is applied to the population of residuals $r_i$, using eight randomly sampled constituents per jet.
We use the Epps--Pulley test statistic to calculate SIGReg, which compares the empirical characteristic function of the residuals with that of a standard normal distribution along random one-dimensional
projections.
We use 1024 random unit vectors and compute the integral from -3 to 3 for each projection using the trapezoidal rule on 17 points.
Subtracting the jet mean prevents a possible failure mode in which every particle token in a jet collapses to the same jet-level vector.

The [CLS] embedding is used directly by the classifier $C$, a single linear layer that maps $z_{[CLS]}$ to class logits.
The classifier is deliberately linear, so that the class information has to be encoded in $z_{[CLS]}$ itself rather than in a large head.
Labels are used only in the cross-entropy loss and never enter the encoder or the predictor.

\subsubsection{Predictor}

The predictor $P$ is a two-block Transformer decoder that predicts the projected embeddings of the hidden constituents using only the context.
The context embeddings $\tilde{z}_i$ of the visible constituents are mapped to the predictor width $d$ by a linear layer, and serve as the keys and values of every cross-attention layer.
The decoder queries are a bank of $N$ learned vectors, initialized from $\mathcal{N}(0, 0.02^2)$ and shared across all jets, and they act as the predictor's working slots.
All meaningful information therefore enters the predictor through cross-attention to the context.
The predictor is not given the positions, features or indices of the target constituents, so it must infer the hidden part of the jet, the target, from the visible part, the context, alone.
Each decoder block applies, in order, self-attention among the queries, cross-attention from the queries to the valid context tokens, and a feed-forward network with hidden width $4d$ and GELU
activation.
Each step has four heads and is wrapped in a pre-layer-normalization residual connection.
A final linear layer maps each query output to the projector dimension $d$, giving a set of candidate latent vectors $\tilde{h}_1, \dots, \tilde{h}_N$.

The predictor outputs an unordered set, while the targets have no natural order either, so the two must be compared with a metric insensitive to permutations.
For a jet with $N_T$ target constituents, the first $N_T$ query outputs form the predicted set.
The pairwise cost is the mean squared distance $c_{ij} = \| \tilde{h}_i - h_j \|^2 / d$ between predicted vector $i$ and target vector $j$.
A soft assignment $\pi_{ij}$ is computed with eight log-domain Sinkhorn iterations at temperature $\tau = 0.1$~\cite{Cuturi:Sinkhorn}.
Tests with small models showed some interesting variability in performance dependent on this choice; our selection of $\tau=0.1$ was made to maximize average performance across the three datasets.
The prediction loss is the transport-weighted cost,
\begin{equation}
\mathcal{L}_{\rm pred} = \frac{1}{N_T} \sum_{i, j} \pi_{ij}\,c_{ij},
\end{equation}
averaged over jets that have a valid target region.
Since $h_j$ is computed by the live encoder and projector, this loss trains the encoder through both the context branch and the target branch.

\subsection{Supervised Classifier}

A supervised classifier is designed to be as close as possible to SuperJEPA.
Its architecture is the same encoder architecture but its [CLS] embedding $z_{[CLS]}$ is passed through two final hidden layers of 256 and 128 units with GELU activations, followed by a linear layer that outputs the class logits.
Its encoder is also trained with dropout of 0.1.

\subsubsection{Model training}

The total SuperJEPA pretraining loss is
\begin{equation}
\mathcal{L}_{\rm total} = \mathcal{L}_{\rm pred} + \lambda\,\mathcal{L}_{\rm SIGReg} + \alpha \,\mathcal{L}_{\rm CE}.
\end{equation}
Our results use $\lambda = 0.01$ which was established as a reasonable setting using runs with small models.
Based on these studies we do not expect our results to be highly sensitive to this choice.
Studies are presented for both $\alpha=0.1$ and $\alpha=1.0$.
The supervised training is nearly identical to SuperJEPA with the prediction and SIGReg losses disabled.

We train two different sized models for each of the supervised, SuperJEPA ($\alpha=0.1$), and SuperJEPA ($\alpha=1$) configurations.
S models have $n=3$ Transformer blocks with $m=4$ heads and $d=128$ while M models have $n=4$ Transformer blocks with $m=8$ heads and $d=256$ (such that head dimension remains fixed).

We train configurations of both S and M models on datasets with sizes 400k, 4M, and in the case of M models also 20M. 
All models are optimized with AdamW, with a global batch size of 256, gradient clipping at a global norm of 1.0 for a fixed number of optimizer steps.
400k models are trained for 100k steps, 4M models are trained for 500k steps, and 20M models are trained for 750k steps.
We observe that the SuperJEPA models are not fully trained after 750k steps and are restarted with the peak learning rate reduced to $1\times10^{-4}$ for another 750k steps.
We observe no improvement from further training on the supervised L model.
In all cases the learning rate rises linearly from zero over the first 8000 steps and then follows a cosine decay to one hundredth of its peak value at the end of training.
The peak base learning rate and weight decay are $10^{-4}$ and $3 \times 10^{-4}$ for both the supervised baseline and SuperJEPA.
We follow the prescription from maximal update parameterization~\cite{muP:TP5} and scale the base learning rate by $1/\sqrt{2}$ for the M models with respect to the S models where the optimal learning rate was selected.
Models are trained on between one and four Nvidia B200s and the learning rate is not scaled with the number of GPUs.
The model is validated every 5000 steps and the weights from the check with the lowest validation loss are kept.
For SuperJEPA, the validation loss is the full pretraining objective, evaluated on target regions that are fixed in advance for the validation set.

After pretraining, the predictor, projector and classifier are discarded.

\subsection{Downstream Evaluation}

During downstream evaluation, we apply the encoder to input jets from JetClass, JetNet, or Top Tagging.
The feature dropout is always disabled so that all available features are used as inputs.
For most studies we freeze the weights of the encoder, but we also consider the option of fine-tuning the encoder weights using the downstream task on the Top Tagging dataset.
The classification in both cases is performed by an MLP classifier.
The classifier is an MLP with two hidden layers of 256 and 128 units and ReLU activations, with dropout of 0.1 after the first hidden layer, and a softmax output over the classes.
When using frozen encoder weights a standard scaler is applied before the MLP and it is trained with categorical cross-entropy loss using the Adam optimizer with an initial learning rate of $2 \times 10^{-4}$ and a batch size of 4096.
The learning rate is reduced by a factor of five if the validation loss does not improve for four epochs.
Training stops after ten epochs without improvement, and the weights with the lowest validation loss are kept.
When training on JetClass, 500k jets from each class are used for a total of 5M.
In the case of JetNet and Top Tagging datasets we use the full available training samples.

In the case of fine-tuning, a classifier head is attached to the [CLS] embedding $z_{[CLS]}$ and training proceeds in two stages.
In the first stage, the encoder is frozen and only the new head is trained for 10 epochs with a learning rate of $3 \times 10^{-4}$.
This prevents large gradients from the randomly initialized head from disturbing the pretrained representation.
In the second stage, all weights are unfrozen and the full network is trained, starting from the best first-stage weights, with a new optimizer and a learning rate reduced by an order of magnitude to $3
\times 10^{-5}$.
Both stages use AdamW with weight decay $1 \times 10^{-4}$, a batch size of 512, gradient clipping at a global norm of 1.0, and the categorical cross-entropy loss.
The learning rate is reduced by a factor of five when the validation loss does not improve for three epochs, and training stops early after seven epochs without improvement.
At the end of each stage, the weights with the lowest validation loss are restored.

\subsubsection{Attention-pooling}

As an alternative to classifying from the [CLS] embedding alone, we also train a classifier on the intermediate embeddings $u_{[CLS]}$ and $u_i$.
The use of these instead of $z_{[CLS]}$ and $z_i$ was found to noticeably improve performance.
The pooling layer has $Q = 4$ learned query vectors that are shared across all jets.
Each query attends to the valid particle tokens of the jet through a single four-head cross-attention layer.
Each query output is projected back to the token width $d$ and layer-normalized.
The $Q$ pooled vectors are concatenated into a single $d_Q = 1024$ and $d_Q = 512$ dimensional jet representation for the M and S models, respectively.
This representation is passed to a two-layer MLP identical to the one described above with a softmax output over the classes.
The pooling layer and the MLP are trained together with the categorical cross-entropy loss, using the Adam optimizer with an initial learning rate of $2 \times 10^{-4}$ and a batch size of 1024.
The optimizer and training settings are otherwise identical to the above classifier training.
Because the pooling has no access to the inter-particle interactions beyond those already encoded by the frozen backbone, any improvement over the [CLS] probe measures how much class information is
present in the particle tokens but not summarized in $z_{[CLS]}$.

\providecommand{\SJstack}[1]{\begin{tabular}[b]{@{}l@{}}SuperJEPA\\($\alpha=#1$)\end{tabular}}

\section{Results}
\label{sec:result}

For all models, we study their performance on downstream tasks using all three datasets described above.
We benchmark their performance by measuring the total area under the curve (AUC) of their receiver operating characteristics as well as their rejection factor for other jets at fixed signal efficiency $X/100$ (Rej$_{X\%}$).

Across tasks we find consistent improvements in performance when training with SuperJEPA compared to an equivalent supervised model (same encoder size, same training data).
In all cases we show comparisons between SuperJEPA ($\alpha=0.1$), SuperJEPA ($\alpha=1$), and an equivalent supervised model.
We summarize our findings below.

\subsection{4-class pretraining}

All 400k and 4M models are pretrained with only four JetClass classes: $H\rightarrow b\bar{b}$, $H\rightarrow gg$, $t\rightarrow bqq'$, and $Z\rightarrow\nu\nu$.
As a test of the ability for these models to adapt to classes outside their pretraining dataset, we consider cases where the downstream task involves all ten JetClass classes.
Table~\ref{tab:jetclass4_rej} shows the performance for all eight model and dataset sizes for each of the three model configurations.
Across model and dataset sizes we observe the best performance across classes from SuperJEPA ($\alpha=1$).
While this remains true even for classes in the pretraining dataset the improvements from SuperJEPA are most substantial in those classes that are very distinct from the pretraining classes, such as $H\rightarrow\ell\nu qq'$ and $t\rightarrow b\ell\nu$.
This is perhaps unsurprising given the lack of jets with leptons in the pretraining, but it is important to note SuperJEPA's ability to obtain separation for these unseen classes.
Some of this separation between supervised and SuperJEPA on unseen classes can be recovered using attention-pooling, but we still observe maximal performance across model and dataset sizes from SuperJEPA.

We also apply these models to the JetNet and Top Tagging datasets and show results in Table~\ref{tab:jetnet_toptag_rej}.
We draw similar conclusions here, where the SuperJEPA ($\alpha=1$) model outperforms the supervised model across model and dataset sizes.
To our knowledge the result for model M (4M) SuperJEPA ($\alpha=1$) with attention-pooling on the Top Tagging dataset represents the best performance in the literature for a frozen encoder.

We also note that for the JetNet dataset, where the jet momentum is significantly different from the JetClass dataset, the SuperJEPA ($\alpha=0.1$) models are in general the best.
This suggests that the exact amount of supervision (i.e. $\alpha)$ that should be used is likely dependent on how many tasks outside of the training phase space the model may be applied to.
One notable exception to SuperJEPA's improved performance over the supervised models is for model M trained on 400k jets.
Here we suspect that the combination of a small pretraining dataset size and large model size results in increased difficulty reaching convergence for SuperJEPA.
We hypothesize that a longer training could alleviate this issue.

\begin{table}[tbp]
\centering
\scriptsize
\begin{tabular}{lcccccccccc}
Configuration & \shortstack{\strut $(*)$ \\ \strut \textbf{$H\to b\bar b$} \\ \strut Rej$_{50\%}$} & \shortstack{\strut \textbf{$H\to c\bar c$} \\ \strut Rej$_{50\%}$} & \shortstack{\strut $(*)$ \\ \strut \textbf{$H\to gg$} \\ \strut Rej$_{50\%}$} & \shortstack{\strut \textbf{$H\to l\nu qq'$} \\ \strut Rej$_{90\%}$} & \shortstack{\strut \textbf{$H\to 4q$} \\ \strut Rej$_{50\%}$} & \shortstack{\strut $(*)$ \\ \strut \textbf{$t\to bqq'$} \\ \strut Rej$_{50\%}$} & \shortstack{\strut \textbf{$t\to bl\nu$} \\ \strut Rej$_{90\%}$} & \shortstack{\strut \textbf{$W\to qq'$} \\ \strut Rej$_{50\%}$} & \shortstack{\strut \textbf{$Z\to q\bar q$} \\ \strut Rej$_{50\%}$} & \shortstack{\strut $(*)$ \\ \strut \textbf{$Z\to\nu\nu$} \\ \strut Rej$_{50\%}$} \\
\midrule
\multicolumn{11}{c}{\textit{No attention pooling}} \\
\midrule
\multicolumn{11}{l}{\underline{Model M (4M)}} \\ [1ex]
\quad Supervised & 966(3) & 135(1) & 106 & 33(1) & 74 & 3634(44) & 227(6) & 62 & 41 & 377(1) \\
\quad \SJstack{0.1} & 393(2) & 86(1) & 98 & \textbf{83(2)} & 93 & 1876(3) & 285(7) & 62 & 38 & 357(1) \\
\quad ($\alpha=1$) & \textbf{1556(5)} & \textbf{231(1)} & \textbf{117} & 72(1) & \textbf{94} & \textbf{4285(26)} & \textbf{384(8)} & \textbf{65} & \textbf{43} & \textbf{388} \\ [1ex]
\multicolumn{11}{l}{\underline{Model M (400k)}} \\ [1ex]
\quad Supervised & 390(3) & 62(1) & 71 & 43(2) & 50 & 1062(14) & 195(8) & \textbf{58} & \textbf{35} & 307(1) \\
\quad \SJstack{0.1} & 211(1) & 45 & 67 & 37(1) & 65 & 788(6) & 173(4) & 51 & 30 & 290(1) \\
\quad ($\alpha=1$) & \textbf{440(2)} & \textbf{73(1)} & \textbf{78} & \textbf{62(2)} & \textbf{66} & \textbf{1374(16)} & \textbf{282(8)} & 54 & 33 & \textbf{311(1)} \\ [1ex]
\multicolumn{11}{l}{\underline{Model S (4M)}} \\ [1ex]
\quad Supervised & 827(1) & 117 & 100 & 24 & 69 & 2589(19) & 150(2) & 64 & 41 & 367(1) \\
\quad \SJstack{0.1} & 530(1) & 86 & 90 & 38 & 77 & 1427(5) & 134(2) & 62 & 38 & 337(1) \\
\quad ($\alpha=1$) & \textbf{965(4)} & \textbf{147(1)} & \textbf{103} & \textbf{46(1)} & \textbf{83} & \textbf{3029(38)} & \textbf{323(6)} & \textbf{66} & \textbf{42} & \textbf{379(1)} \\ [1ex]
\multicolumn{11}{l}{\underline{Model S (400k)}} \\ [1ex]
\quad Supervised & \textbf{421(2)} & \textbf{69} & 72 & 40 & 53 & 1011(6) & 134(1) & \textbf{57} & \textbf{35} & 307(1) \\
\quad \SJstack{0.1} & 208(1) & 43 & 62 & \textbf{48} & 55 & 503(5) & 125(3) & 55 & 32 & 266(1) \\
\quad ($\alpha=1$) & 397(1) & 68(1) & \textbf{74} & 37(1) & \textbf{56} & \textbf{1072(6)} & \textbf{219(3)} & 56 & 34 & \textbf{316(2)} \\
\midrule
\multicolumn{11}{c}{\textit{Attention pooling}} \\
\midrule
\multicolumn{11}{l}{\underline{Model M (4M)}} \\ [1ex]
\quad Supervised & 1390(13) & 237(3) & 124(1) & 630(27) & 154(3) & 4914(9) & 1779(84) & 82(1) & 54(1) & 414(1) \\
\quad \SJstack{0.1} & 547(19) & 116(3) & 109(1) & 370(53) & 135(4) & 2268(39) & 929(97) & 72(1) & 46(1) & 379(3) \\
\quad ($\alpha=1$) & \textbf{1951(4)} & \textbf{330(5)} & \textbf{135(1)} & \textbf{666(36)} & \textbf{170(3)} & \textbf{5802(76)} & \textbf{1932(102)} & \textbf{83(1)} & \textbf{55(1)} & \textbf{420(3)} \\ [1ex]
\multicolumn{11}{l}{\underline{Model M (400k)}} \\ [1ex]
\quad Supervised & 630(5) & 117(1) & 88 & 281(6) & 89(1) & 1616(19) & 864(5) & \textbf{72} & \textbf{45} & 339(1) \\
\quad \SJstack{0.1} & 334(13) & 69(2) & 82(1) & 234(25) & 110(2) & 1155(20) & 637(34) & 64(1) & 39 & 330(2) \\
\quad ($\alpha=1$) & \textbf{720(4)} & \textbf{133(1)} & \textbf{102(1)} & \textbf{364(8)} & \textbf{121(1)} & \textbf{2046(27)} & \textbf{1009(18)} & \textbf{72} & \textbf{45} & \textbf{367(1)} \\ [1ex]
\multicolumn{11}{l}{\underline{Model S (4M)}} \\ [1ex]
\quad Supervised & 1140(18) & 193(3) & 114(1) & \textbf{460(29)} & 127(4) & \textbf{3769(104)} & \textbf{1408(76)} & 78(1) & 51(1) & \textbf{400(2)} \\
\quad \SJstack{0.1} & 729(34) & 130(7) & 100(1) & 268(40) & 113(5) & 1917(69) & 765(99) & 71(1) & 45(1) & 361(4) \\
\quad ($\alpha=1$) & \textbf{1283(15)} & \textbf{213(6)} & \textbf{117(1)} & 395(35) & \textbf{134(3)} & \textbf{3744(36)} & 1303(79) & \textbf{79} & \textbf{52(1)} & 398(2) \\ [1ex]
\multicolumn{11}{l}{\underline{Model S (400k)}} \\ [1ex]
\quad Supervised & \textbf{596(13)} & \textbf{111(2)} & 85(1) & \textbf{234(19)} & 83(2) & \textbf{1498(32)} & \textbf{730(51)} & \textbf{70(1)} & \textbf{44} & 337(2) \\
\quad \SJstack{0.1} & 272(7) & 57(1) & 71(1) & 132(6) & 80(2) & 730(25) & 424(19) & 62(1) & 37 & 289(1) \\
\quad ($\alpha=1$) & \textbf{611(17)} & \textbf{114(4)} & \textbf{89(1)} & \textbf{226(19)} & \textbf{91(3)} & \textbf{1498(52)} & \textbf{747(52)} & \textbf{70(1)} & 43(1) & \textbf{349(3)} \\
\bottomrule
\end{tabular}
\caption{Results on the JetClass dataset for supervised and SuperJEPA models pretrained only on the 4 classes denoted with $(*)$. Numbers in parentheses next to models represent the size of the JetClass pretraining dataset. Numbers in parentheses next to results represent uncertainty from three classifier training runs. Downstream models are shown both with and without attention pooling.}\label{tab:jetclass4_rej}
\end{table}

\begin{table}[tbp]
\centering
\footnotesize
\begin{tabular}{lccccc!{\hspace{3pt}\vrule\hspace{3pt}}cccc}
 & \multicolumn{5}{c}{\textbf{JetNet}} & \multicolumn{4}{c}{\textbf{Top Tagging}} \\ \cmidrule(lr){2-6} \cmidrule(lr){7-10}
 & \multicolumn{7}{c}{\textit{No attention pooling}} & \multicolumn{2}{c}{\textit{Attention pooling}} \\ \cmidrule(lr){2-8} \cmidrule(lr){9-10}
Configuration & \shortstack{\strut \textbf{$q$} \\ \strut Rej$_{50\%}$} & \shortstack{\strut \textbf{$g$} \\ \strut Rej$_{50\%}$} & \shortstack{\strut \textbf{$w$} \\ \strut Rej$_{50\%}$} & \shortstack{\strut \textbf{$z$} \\ \strut Rej$_{50\%}$} & \shortstack{\strut \textbf{$t$} \\ \strut Rej$_{50\%}$} & \shortstack{\strut  \\ \strut Rej$_{50\%}$} & \shortstack{\strut  \\ \strut Rej$_{30\%}$} & \shortstack{\strut  \\ \strut Rej$_{50\%}$} & \shortstack{\strut  \\ \strut Rej$_{30\%}$} \\
\midrule
\multicolumn{6}{l!{\hspace{3pt}\vrule\hspace{3pt}}}{\underline{Model M (4M)}} & \multicolumn{4}{c}{} \\ [1ex]
\quad Supervised & \textbf{38} & 35 & 117(1) & 261(3) & 168(1) & 344(3) & 1256(27) & 391(5) & 1416(49) \\
\quad \SJstack{0.1} & \textbf{38} & 37 & \textbf{120(2)} & \textbf{291(8)} & \textbf{201(2)} & 326(6) & 1239(50) & 353(7) & 1342(68) \\
\quad ($\alpha=1$) & \textbf{38} & \textbf{39} & 95(1) & 247(2) & \textbf{197(6)} & \textbf{365(8)} & \textbf{1540(25)} & \textbf{403(9)} & \textbf{1883(10)} \\ [1ex]
\multicolumn{6}{l!{\hspace{3pt}\vrule\hspace{3pt}}}{\underline{Model M (400k)}} & \multicolumn{4}{c}{} \\ [1ex]
\quad Supervised & \textbf{37} & \textbf{35} & \textbf{163(2)} & \textbf{395(8)} & 154(3) & 211(2) & 801(38) & 251(7) & 941(33) \\
\quad \SJstack{0.1} & \textbf{37} & 33 & 87(1) & 183(3) & \textbf{167(3)} & 250(6) & 916(33) & 286(3) & \textbf{1167(8)} \\
\quad ($\alpha=1$) & \textbf{37} & 34 & 94(1) & 212(3) & 161(2) & \textbf{263(5)} & \textbf{976(16)} & \textbf{303(3)} & \textbf{1189(106)} \\ [1ex]
\multicolumn{6}{l!{\hspace{3pt}\vrule\hspace{3pt}}}{\underline{Model S (4M)}} & \multicolumn{4}{c}{} \\ [1ex]
\quad Supervised & \textbf{38} & 35 & 129(1) & 298(3) & 171(1) & \textbf{327(5)} & \textbf{1290(33)} & 339(5) & \textbf{1472(40)} \\
\quad \SJstack{0.1} & \textbf{38} & \textbf{37} & \textbf{141(2)} & \textbf{351(7)} & 189(3) & 281(1) & 1192(26) & 306(1) & 1202(41) \\
\quad ($\alpha=1$) & 37 & 35 & 122(2) & 322(1) & \textbf{195(1)} & \textbf{327(3)} & \textbf{1277(39)} & \textbf{354(7)} & \textbf{1423(61)} \\ [1ex]
\multicolumn{6}{l!{\hspace{3pt}\vrule\hspace{3pt}}}{\underline{Model S (400k)}} & \multicolumn{4}{c}{} \\ [1ex]
\quad Supervised & \textbf{37} & \textbf{36} & 147(1) & 340(5) & 141(4) & 215(3) & 797(11) & 237(4) & 878(13) \\
\quad \SJstack{0.1} & \textbf{37} & \textbf{36} & \textbf{160(2)} & \textbf{361(6)} & 158(1) & 225(5) & 833(67) & 236(3) & 919(2) \\
\quad ($\alpha=1$) & \textbf{37} & \textbf{36} & 142(2) & 341(7) & \textbf{169(2)} & \textbf{231(2)} & \textbf{901(24)} & \textbf{265(1)} & \textbf{1028(36)} \\
\bottomrule
\end{tabular}
\caption{Results on the JetNet and Top tagging datasets for supervised and SuperJEPA models pretrained only on 4 classes from JetClass. Numbers in parentheses next to models represent the size of the JetClass pretraining dataset. Numbers in parentheses next to results represent uncertainty from three classifier training runs. Downstream models are shown both with and without attention pooling.}\label{tab:jetnet_toptag_rej}
\end{table}

\subsection{10-class pretraining}

To study the performance of SuperJEPA on larger pretraining datasets, we pretrain model M using 2M jets from all ten classes for a total of 20M jets.
The results of these models on JetClass with and without attention-pooling are shown in Table~\ref{tab:jetclass10_rej}.
We again observe the best performance across classes from the SuperJEPA ($\alpha=1$) model.
This remains true both with and without attention-pooling, although, unsurprisingly, we observe minimal improvement in this case given that the pretraining dataset matches exactly to the downstream task.

\begin{table}[tbp]
\centering
\scriptsize
\begin{tabular}{lcccccccccc}
Configuration & \shortstack{\strut \textbf{$H\to b\bar b$} \\ \strut Rej$_{50\%}$} & \shortstack{\strut \textbf{$H\to c\bar c$} \\ \strut Rej$_{50\%}$} & \shortstack{\strut \textbf{$H\to gg$} \\ \strut Rej$_{50\%}$} & \shortstack{\strut \textbf{$H\to l\nu qq'$} \\ \strut Rej$_{90\%}$} & \shortstack{\strut \textbf{$H\to 4q$} \\ \strut Rej$_{50\%}$} & \shortstack{\strut \textbf{$t\to bqq'$} \\ \strut Rej$_{50\%}$} & \shortstack{\strut \textbf{$t\to bl\nu$} \\ \strut Rej$_{90\%}$} & \shortstack{\strut \textbf{$W\to qq'$} \\ \strut Rej$_{50\%}$} & \shortstack{\strut \textbf{$Z\to q\bar q$} \\ \strut Rej$_{50\%}$} & \shortstack{\strut \textbf{$Z\to\nu\nu$} \\ \strut Rej$_{50\%}$} \\
\midrule
\multicolumn{11}{c}{\textit{No attention pooling}} \\
\midrule
\quad Supervised & 2360(3) & 424(1) & 150 & 1444(3) & 235 & 7772(31) & 4749(23) & 93 & 65 & 449(1) \\
\quad \SJstack{0.1} & 1776(2) & 311 & 135 & 995(2) & 195 & 4398(7) & 2709(18) & 86 & 58 & 423(1) \\
\quad ($\alpha=1$) & \textbf{2530(3)} & \textbf{458(2)} & \textbf{153} & \textbf{1508(3)} & \textbf{242} & \textbf{8461(77)} & \textbf{5079(4)} & \textbf{95} & \textbf{66} & \textbf{451} \\
\midrule
\multicolumn{11}{c}{\textit{Attention pooling}} \\
\midrule
\quad Supervised & 2349(22) & 423(1) & 150 & 1431(11) & 233(1) & 7643(60) & 4697(17) & 93 & 64 & 448(1) \\
\quad \SJstack{0.1} & 1800(4) & 315(1) & 137 & 1004(7) & 197 & 4555(23) & 2772(7) & 87 & 58 & 424(1) \\
\quad ($\alpha=1$) & \textbf{2519(7)} & \textbf{458(1)} & \textbf{153} & \textbf{1494(2)} & \textbf{240} & \textbf{8317(57)} & \textbf{4985(21)} & \textbf{95} & \textbf{66} & \textbf{451(2)} \\
\bottomrule
\end{tabular}
\caption{Results on the JetClass dataset for supervised and SuperJEPA models pretrained on 20M total samples from all 10 classes. Numbers in parentheses next to models represent the size of the JetClass pretraining dataset. Numbers in parentheses next to results represent uncertainty from three classifier training runs. Downstream models are shown both with and without attention pooling.}\label{tab:jetclass10_rej}
\end{table}

\subsection{Fine-tuning}

We also apply all configurations of model M pretrained with 400k, 4M, and 20M jets to the Top Tagging dataset.
This time we allow the encoder weights to train instead of leaving them frozen.
The results are summarized in Table~\ref{tab:toptag_finetune}.
We observe the most competition in these conditions between the supervised and SuperJEPA ($\alpha=1$) models.
However, SuperJEPA still achieves the best overall result across all metrics we consider for each dataset size.

\begin{table}[tbp]
\centering
\footnotesize
\begin{tabular}{lccc}
Configuration & AUC & Rej$_{50\%}$ & Rej$_{30\%}$ \\
\midrule
\multicolumn{4}{l}{\underline{Model M (20M, 10 classes)}} \\ [1ex]
\quad Supervised & \textbf{0.9861(1)} & \textbf{474.4(4.3)} & 1915(29) \\
\quad \SJstack{0.1} & 0.9858 & 408.8(3.5) & 1798(16) \\
\quad ($\alpha=1$) & \textbf{0.9862(1)} & \textbf{470.9(5.4)} & \textbf{1980(46)} \\ [1ex]
\multicolumn{4}{l}{\underline{Model M (4M, 4 classes)}} \\ [1ex]
\quad Supervised & 0.9857 & 416.8(5.6) & 1524(52) \\
\quad \SJstack{0.1} & 0.9855(1) & 399.4(1.7) & 1615(67) \\
\quad ($\alpha=1$) & \textbf{0.9861} & \textbf{449.7(2.2)} & \textbf{1769(55)}  \\ [1ex]
\multicolumn{4}{l}{\underline{Model S (400k, 4 classes)}} \\ [1ex]
\quad Supervised & 0.9839(1) & \textbf{337.8(2.8)} & \textbf{1282(44)} \\
\quad \SJstack{0.1} & 0.9839(1) & 307.5(1.6) & 1239(27)  \\
\quad ($\alpha=1$) & \textbf{0.9843} & \textbf{334.1(6.0)} & \textbf{1313(15)} \\ [1ex]
\bottomrule
\end{tabular}
\caption{Results after fine-tuning on the top tagging dataset for supervised and SuperJEPA models. Numbers in parentheses next to models represent the size of the JetClass pretraining dataset. Numbers in parentheses next to results represent uncertainty from three finetuning runs.}\label{tab:toptag_finetune}
\end{table}

\subsection{Representations}

The preceding results establish that SuperJEPA transfers better to a variety of downstream tasks.
We next consider whether this improvement is accompanied by visible differences in the learned representation, or whether the changes are more subtle.

\begin{figure}[t]
  \centering
  \begin{subfigure}{0.45\textwidth}
    \includegraphics[width=\linewidth]{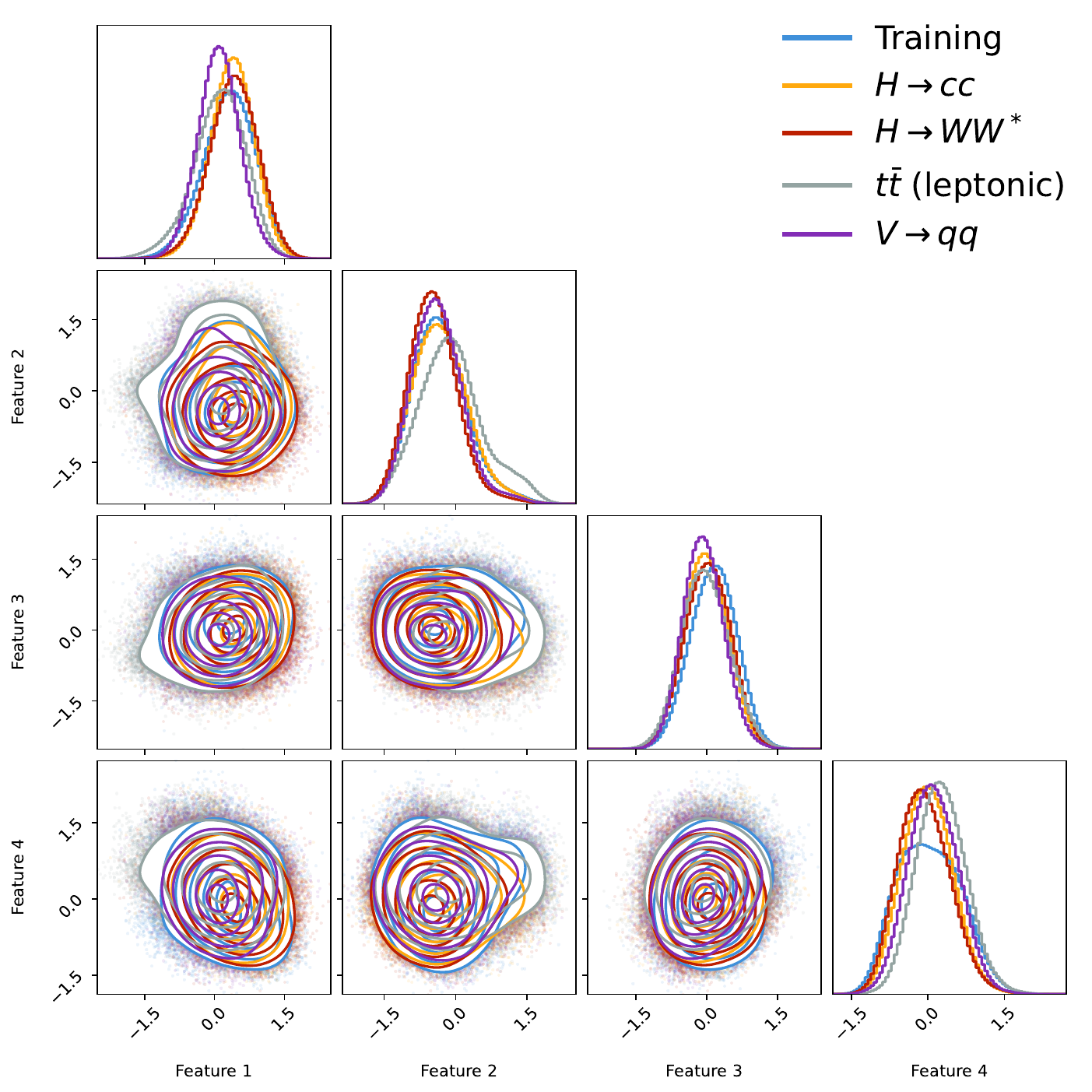}
    \caption{Supervised.}
    \label{fig:partfeats_supervised}
  \end{subfigure}
  \hfill
  \begin{subfigure}{0.45\textwidth}
    \includegraphics[width=\linewidth]{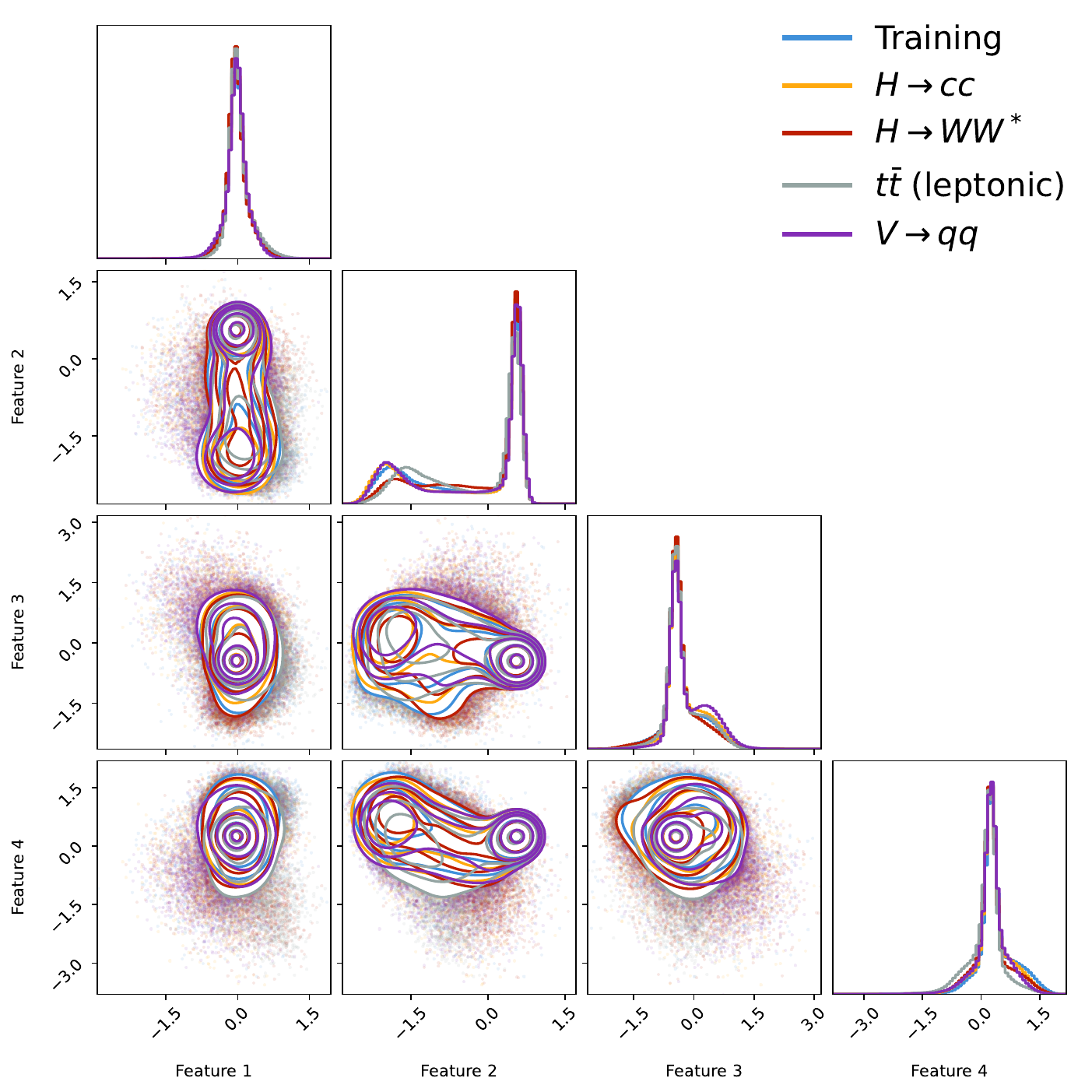}
    \caption{SuperJEPA ($\alpha=1$).}
    \label{fig:partfeats_superjepa_alpha1p0}
  \end{subfigure}
  \caption{Corner plots of four per-particle embedding features as produced by an encoder trained using only standard supervision (a) and SuperJEPA with $\alpha=1$ (b). The distribution shown in blue represents classes seen during pretraining, while the other colors represent new classes not used in pretraining.}
  \label{fig:partfeats}
\end{figure}

In order to study the representations produced by different models, we measure the distributions of particle embeddings, $z_i$, from the models.
Fig.~\ref{fig:partfeats} shows the first few dimensions of the particle embeddings $z_i$ from the model M supervised and SuperJEPA ($\alpha=1$) models pretrained with 4M jets.
We observed reasonably Gaussian distributions of these embeddings from the supervised model and not SuperJEPA.
This occurs in spite of the SIGReg loss term in SuperJEPA (and lack of any distribution pressure in the supervised pretraining).
While this behavior deserves additional study, we note that the SIGReg loss term serves only to prevent model collapse during training. 
Given the prediction task's need to allow for prediction of the rest of the target particles, the particle embeddings in SuperJEPA must retain significant information.
On the other hand, supervised trainings have no need to retain any information in the particle embeddings and therefore produce substantially less structured class-dependent variation.
Notably, we observe minimal difference in the particle embeddings for classes seen during pretraining vs. not seen during pretraining.

We also study the class embeddings $z_{[CLS]}$ produced by encoders of different sizes and training configurations. 
We use t-SNE~\cite{tSNE} to reduce the embeddings to two dimensions for plotting purposes.
Unsurprisingly, we see improved class separation as model and dataset sizes increase.
In all cases, the largest models produce clearly useful representations for class separation regardless of training scheme.
We find little visual difference in the representations from supervised or SuperJEPA encoders.

\begin{figure}
    \centering
    \includegraphics[width=0.75\linewidth]{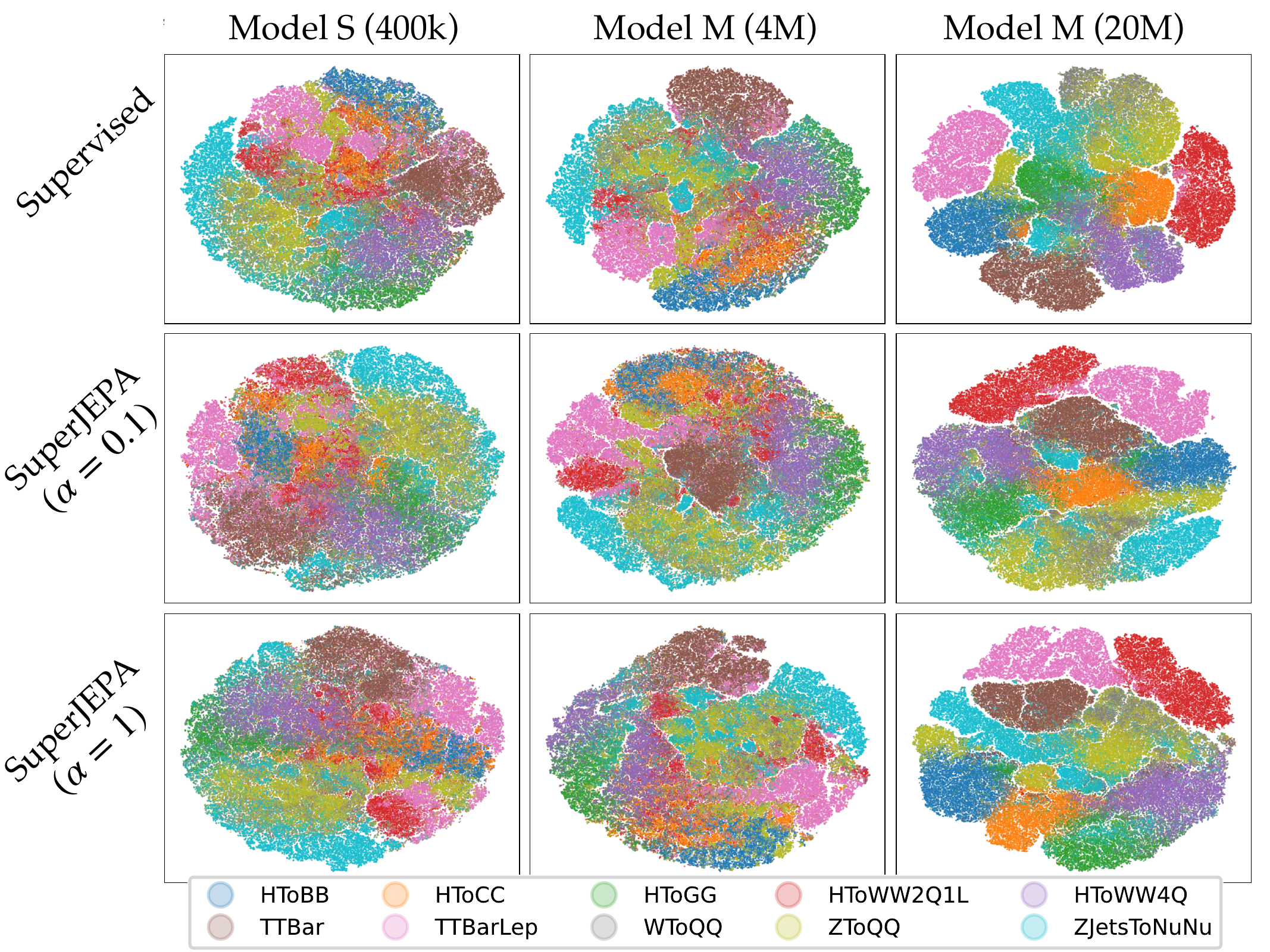}
    \caption{t-SNE plots of the CLS token for different models. Supervised models are shown in the top row, SuperJEPA $(\alpha=0.1)$ models are shown in the middle row, and SuperJEPA $(\alpha=1)$ models are shown in the bottom row. Model and pretraining dataset size increase from left to right. All JetClass classes are shown.}
    \label{fig:tsnes}
\end{figure}

\subsection{Ablations}

There are many components in the SuperJEPA workflow that can meaningfully affect performance.
We quantify the impact of individual design choices by retraining model S on the 400k-jet dataset with a single modification at a time and quoting the relative decrease in background rejection with respect to the nominal SuperJEPA configuration with the same $\alpha$.
Rejections are evaluated at a signal efficiency of 50\% (90\% for $H\to l\nu qq'$ and $t\to bl\nu$) and averaged geometrically over classes, so that negative values indicate an improvement over the nominal.
The training variants are then compared to the seed-averaged nominal we use above, whose seed-to-seed spread of at most 3\% sets the scale for a significant change.

\paragraph{Masking scheme.} For $\alpha=0.1$ ($\alpha=1$), replacing the nominal masking scheme based on subjets with an entirely random one reduces the rejection by 5--18\% (5--23\%) on JetClass and 56\%(25--26\%) for top tagging.
In this test rejection on JetNet is reduced by 11\% with $\alpha=0.1$ while no meaningful change is observed for $\alpha=1$.
We conclude from these tests that the structured SuperJEPA masking scheme provides a significant improvement with respect to a random masking scheme across almost all datasets and $\alpha$ values.

\paragraph{Sinkhorn matching.} Lowering the Sinkhorn temperature to $\tau=0.01$ with $\alpha=0.1$ improves the rejection on all three tasks, by 25--28\% on JetClass, 10\% on JetNet and 5--15\% for top tagging, while tests with $\alpha=1$ show no significant effect beyond a 9\% decrease of the JetClass rejection without attention pooling.
Raising the Sinkhorn temperature to $\tau=1$ with $\alpha=0.1$ increases the JetClass rejection by 18--44\% and the top-tagging rejection by 8--10\%, but reduces the JetNet rejection by 19\%.
The same tests with $\alpha=1$ show increases in JetClass rejection by 8--12\% and slight improvements to top tagging, but a reduction in the JetNet rejection by 25\%.

\paragraph{Subjet construction.} We observe that adjusting the preprocessing to create four subjets changes the JetClass rejection by at most 10\%, while increasing to 16 subjets or adjusting the clustering to use exclusive-$k_T$ reduce it by 18--36\% and also degrade JetNet performance.

\paragraph{SIGReg.} The use of a CLS regularization based on SIGReg was found to decrease performance by 4--6\% on JetClass and 13\% on JetNet, while adjustments of $\lambda$ were observed to have minimal or negative effects of up to 15\%, suggesting the default represents an optimal choice already.
\section{Discussion}
\label{sec:discussion}

Our results demonstrate improved performance of SuperJEPA over supervised models across a range of model sizes and tasks.
Crucially, these results also appear to hold even as the models are exposed to increasing dataset sizes.
While we use pretraining datasets with up to 20M jets, we are unable to test on datasets into the 100Ms and 1Bs.
Future studies are necessary to understand if our conclusions continue to hold with these even larger datasets.
In this work we have also considered both frozen and fine-tuned encoders; we expect that the need for one over the other is likely to rely on the fine-tuning cost, model size, and downstream dataset size.
A one size fits all solution seems unlikely, and we therefore expect that performance in both situations is likely to be important for any potential HEP foundation model.

While we have considered only the use of fully labeled data in this work, we note that our method could be easily adapted to use unlabeled data as well.
A simple modification to allow this would be to remove the classification loss for these events, but other methods may also exist to help improve the performance on these unlabeled samples.
We also expect that the optimal settings for SuperJEPA might change meaningfully in the presence of substantial unlabeled data.

The observed preference for strong supervision ($\alpha=1$) with SuperJEPA is consistent with recent observations that supervised pretraining remains particularly effective when accurate labels are abundant~\cite{Elsharkawy:objectives}. 
Our results suggest that predictive self-supervision can nevertheless provide additional benefit when it supplements this objective, rather than replacing or substantially down-weighting it.
However, the balance between these objectives is clearly dataset dependent, and we find that when transferring to JetNet weaker supervision is preferred.

Finally, we note that our work, and indeed much of the previous work with SSL in collider physics, has focused on jet-tagging.
The application of SSL techniques, like SuperJEPA, for alternative tasks like particle identification or event classification is left for future work.

\section{Conclusion}
\label{sec:conclusion}

SuperJEPA combines a joint-embedding predictive objective with supervised pretraining and improves downstream performance relative to supervised training alone across the jet-tagging studies considered here. The improvement is observed for both frozen and fine-tuned encoders and across multiple model and pretraining-dataset sizes. These results indicate that predictive self-supervision can remain useful even when large labeled simulation samples are available, and motivate further tests at larger scales, with partially unlabeled data, and on particle-physics tasks beyond jet tagging.

\section*{Acknowledgements}

\paragraph{Funding information}
HFT and DSR are supported by the U.S. Department of Energy (DOE), Office of Science, Office of High Energy Physics subprogram on Computational High Energy Physics under Award No. DE-SC0026801. 
DSR is also supported by the U.S. Department of Energy (DOE), Office of Science, Office of High Energy Physics Early Career Research program under Award No. DE-SC0025324.

\bibliography{references}

\end{document}